\documentclass[aps,prl,twocolumn,groupedaddress,showpacs]{revtex4}

\usepackage{graphicx}
\usepackage{amsmath}

\begin{document}

%\preprint{}

\title{Vortex phase diagram in rotating two-component Bose-Einstein condensates}


\author{Kenichi Kasamatsu$^1$}
\author{Makoto Tsubota$^1$}
\author{Masahito Ueda$^2$}
%\thanks{}

\affiliation{
$^1$Department of Physics,
Osaka City University, Sumiyoshi-Ku, Osaka 558-8585, Japan 
\\
$^2$Department of Physics, Tokyo Institute of Technology,  
Meguro-ku, Tokyo 152-8551, Japan}


\date{\today}

\begin{abstract}
We investigate the structure of vortex states 
in rotating two-component Bose-Einstein condensates with 
equal intracomponent but varying intercomponent coupling constants. 
A phase diagram in the intercomponent-coupling versus rotation-frequency 
plane reveals rich equilibrium structures of vortex states. 
As the ratio of intercomponent to intracomponent couplings increases, 
the interlocked vortex lattices undergo phase transitions from triangular to square, 
to double-core lattices, and eventually develop interwoven ``serpentine" vortex sheets 
with each component made up of chains of singly quantized vortices. 
\end{abstract}

%\keywords{}

\maketitle

A system described by a multicomponent order parameters can excite various exotic topological defects. 
There have been great interest in the study of such topological defects, 
which has been explored extensivly in condensed matter systems such as anisotropic 
superfluid $^{3}$He \cite{helium3}. This study is also closely related with some problems 
in superconductors of UPt$_{3}$ and Sr$_{2}$RuO$_{4}$ which are thought to have 
non-s-wave symmetry \cite{supercond} and high-energy physics and cosmology \cite{vorbook}. 
Dilute atomic Bose-Einstein condensates (BECs) offer another superior testing ground to investigate 
the physics of exotic topological defects because they are free from impurities and well controlled by optical techniques. 
A trapped BEC in a rotating potential generates quantized vortices which are topological defects 
characteristic of superfluidity, and several groups have suceeded in observing ordered vortex lattices \cite{Madison}. 
While many interesting phenomena have been found in single-component BECs \cite{Madison},
a rich variety of static and dynamic phenomena are expected to 
occur in a system of rotating two-component BECs consisting, for example,
of two different hyperfine spin states of atoms \cite{Hall}. 
This system is characterized by three coupling constants denoted by $C_{11}$ and 
$C_{22}$ (for intracomponents) and $C_{12}$ (for intercomponent) \cite{Ho,Ersy,Ohberg1,Pu,Tim}. 
In this Letter, we report equilibrium structures of 
vortex states in two-component BECs with $C_{11}=C_{22} \equiv C$ but varying 
values of $C_{12}$ and the rotation frequency based 
on the numerical analysis of the Gross-Pitaevskii equation. 

Mueller and Ho studied the vortex lattice structure of two-component BECs 
by assuming the lowest Landau level approximation and 
a perfect lattice \cite{Mueller}. They proposed a phase diagram 
on the lattice structure and found that
as $C_{12}/C$ is increased, the system undergoes a continuous structural 
change from triangular to square lattices. 
Kita {\it et al.} discussed the lattice structure in $F=1$ spinor BECs \cite{Kita}.
However, both studies assume the fast rotation limit where the rotation frequency is very close to the radial trapping 
frequency, so a vast experimentally accessible parameter regime remains to be explored. 
When there is no rotation, it is known that we obtain miscible condensates 
as the ground state for $C_{12}/C <1$, 
while for $C_{12}/C >1$ the two components phase separate \cite{Ho,Ersy,Tim}. 
In the presence of rotation, our numerical analysis reveals a rich variety of vortex states which have 
eluded analytic treatments. In particular, we find that for $C_{12}/C \geq 1$ double-core vortex 
lattices or interwoven ``serpentine" vortex sheets become stabilized for a considerable range 
of rotation frequency; these states can therefore be observed in current experimental situations. 

The equilibrium solutions of two-component BECs in a frame rotating 
with the angular velocity ${\bf \Omega}=\Omega {\bf \hat{z}}$
are obtained by solving the time-independent coupled Gross-Pitaevskii (GP) equations 
for the condensate wave functions $( h_{i} + \sum_{j=1,2} C_{ij}|\Psi_{j}|^{2} ) 
\Psi_{i} = \mu_{i} \Psi_{i}$ ($i=1,2$) \cite{Ho,Ersy,Ohberg1,Pu,Tim}, 
where $h_{i}=- (m_{12}/m_{i}) \nabla^{2} + V_{i} - (\Omega/\bar{\omega}) L_{z}$ 
with the reduced mass $m_{12}=m_{1}m_{2}/(m_{1}+m_{2})$ and 
the harmonic potential $V_{i}=(m_{i} \omega_{i}^{2}/4 m_{12} \bar{\omega}^{2}) r^{2}$ . 
Here, length and energy are measured in units of 
$b_{\rm ho}=\sqrt{\hbar/2 m_{12} \bar{\omega}}$ 
and $\hbar \bar{\omega} \equiv \hbar (\omega_{1} + \omega_{2})/2$, respectively. 
In this work, the numerical calculation is done in the two-dimensional $x-y$ space. 
The wave functions are normalized as $\int dx dy |\Psi_{i}(x,y)|^{2} = 1$ and, then, 
the intra- and intercomponent coupling constants are written 
as $C_{ii}=8 \pi (m_{12}/m_{i}) N_{i}^{\rm 2d} a_{ii}$ 
and $C_{ij}=4 \pi N_{j}^{\rm 2d} a_{12}$ $(i \neq j)$
with the corresponding s-wave scattering lengths $a_{11}$, $a_{22}$, $a_{12}$ 
(assumed to be positive), and the particle numbers $N_{i}^{\rm 2d}$ per unit length along the $z$-axis. 

For simplicity, the number of the parameters is reduced 
by assuming $a_{11}=a_{22}$, $m_{1}=m_{2}$ and $\omega_{1}=\omega_{2}=\bar{\omega}$, and
$N_{1}^{\rm 2d}=N_{2}^{\rm 2d}=N^{\rm 2d}$. 
For the typical experimental conditions 
$( \omega_{\perp}, \omega_{z} )=2\pi ( 8, 5 )$Hz and $N \sim 10^{6}$, 
the values of the intracomponent coupling constants are chosen as $C_{11} = C_{22} = C = 2000$. 
In the numerical calculation, the equilibrium solutions are found by the norm-preserving imaginary 
time propagation of the time-dependent coupled GP equations, starting from arbitrary trial 
wave functions without vorticity. The propagation continues until the fluctuation in the 
norm of the wave function becomes smaller than 10$^{-8}$. 

\begin{figure}[btp]
\includegraphics[height=0.26\textheight]{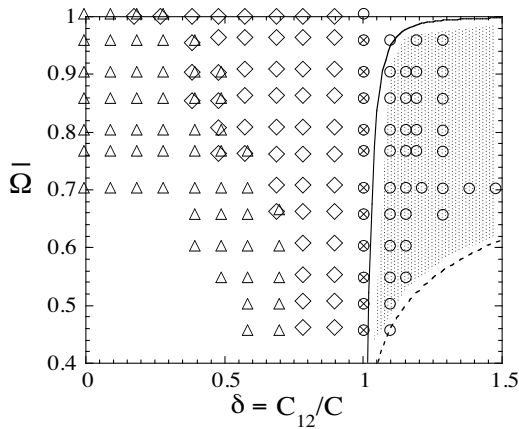}
\caption{$\bar{\Omega}$-$\delta$ phase diagram of the vortex states 
in rotating two-component BECs. The classification of the structure is made by the symbols,
$\triangle$: triangular lattice, $\diamond$: square lattice, 
$\otimes$: stripe or double-core vortex lattice, $\circ$: vortex sheet. 
Because of the continuous change from triangular to square, 
their boundary is shown by using both $\triangle$ and $\diamond$. 
The plots at $\bar{\Omega}=1$ show the results of Ref. \cite{Mueller}. 
The shaded region shows the analytically obtained region of vortex sheets;  
the solid curve represents $b=\Lambda_{p}$ and the dashed curve 
$2b=R_{\rm TF}$  (see text). }
\label{phasedia}
\end{figure}

With a fixed value $C = 2000$, 
we investigate the equilibrium solutions for various values of two free parameters 
$\delta \equiv C_{12}/C$ and $\bar{\Omega} \equiv \Omega/\bar{\omega}$. 
The obtained vortex structure is summarized in the phase diagram of Fig. \ref{phasedia}. 
Because the centrifugal potential overcomes the harmonic trapping potential
for $\bar{\Omega}>1$, the upper limit of the rotation frequency is set by $\bar{\Omega}=1$. 
Previous studies have revealed that two non-rotating components 
with equal mass and equal intracomponent scattering length overlap completely 
for $\delta<1$, and phase-separate for $\delta>1$ \cite{Ho,Ersy,Tim}. 
This criterion is reflected on the structure of vortex states, as explained below. 
In general, there are many metastable states in rotating two-component BECs; 
they are obtained by the numerical simulation starting from different wave functions. 
Here we discuss some characteristic features of vortex structures. 

\begin{figure}[btp]
\includegraphics[height=0.43\textheight]{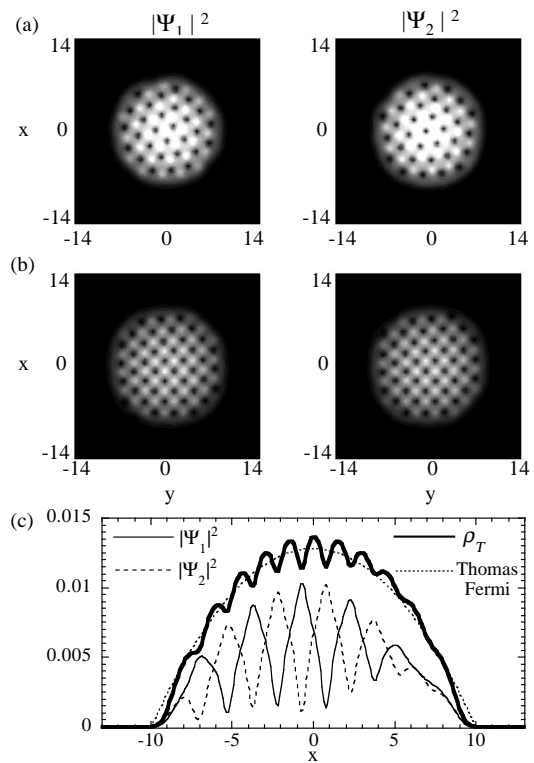}
\caption{Two-dimensional density profile of the condensates with $C=2000$ and $\delta=0.7$ 
for (a) $\bar{\Omega}=0.6$ and (b) $\bar{\Omega}=0.75$. (c) Cross section of (b) along the $y=0$ line. 
The total density $\rho_{T}=|\Psi_{1}|^{2}+|\Psi_{2}|^{2}$ and the Thomas-Fermi density profile are also shown.}
\label{vortexlattice}
\end{figure}

In the region $\delta<1$, two types of the regular vortex lattices 
are obtained as the equilibrium state. For $\delta=0$ where two components 
are not interacting, the formulation is equivalent to that of one component, 
thereby triangular vortex lattices are formed \cite{Tsubota}.  
As $\delta$ increases, the positions of vortex cores in one component gradually 
shift from those of the other component and the triangular lattices are distorted. 
Eventually, the vortices in each component 
form a square lattice rather than a triangular one. 
The $\delta$ dependence on the stable region of a square lattice was studied 
in Ref. \cite{Mueller} in the high rotation limit $\bar{\Omega} \simeq 1$. 
As seen in Fig. \ref{phasedia}, however, that stable region 
depends not only on $\delta$ but also sensitively on $\bar{\Omega}$. 
An increase in rotation frequency also causes the transition from triangular 
to square lattices; in Figs. \ref{vortexlattice} (a) and (b), 
the two-dimensional density profiles of the condensates with $\delta=0.7$ 
are shown for $\bar{\Omega}=0.6$ and $\bar{\Omega}=0.75$. 
We find that the two vortex lattices are interlocked in such a manner that a 
peak in the density of one component is located at the density hole of the other, 
as shown in Fig. \ref{vortexlattice} (c). 
As a result, the total density $\rho_{T}= |\Psi_{1}|^{2} + |\Psi_{2}|^{2}$ obeys 
the Thomas-Fermi distribution applied to the overlapping two-component BECs 
with solid-body rotation $\rho_{T}(r) = 2 \sqrt{\alpha/\pi} -\alpha r^{2}$ 
with $\alpha=(1-\bar{\Omega}^{2})/C(1+\delta)$. 
We have confirmed that this feature holds for other parameter regimes. 

Why is the square lattice stabilized in two-component BECs? 
According to the energy functional of this system, two components interact 
via the intercomponent interaction $C_{12} |\Psi_{1}|^{2} |\Psi_{2}|^{2}$; 
the velocity field in one component is independent of that of the other. 
Therefore, such a feature is determined {\it only by the density distribution of the condensates} 
which minimizes the interaction energy. 
We rewrite the interaction energy in terms of the total density $\rho_{T}$ 
and the ``spin" variable $S = |\Psi_{1}|^{2} - |\Psi_{2}|^{2}$ as 
%\begin{equation}
$E_{\rm int} = \int d {\bf r} C [ (1 +\delta) \rho_{T}^{2} + (1 - \delta) S^{2}] /4$. 
%\label{intenergy}
%\end{equation} 
A larger $\delta$ makes a smoother total density $\rho_{T}$ more favorable in order to reduce 
the first term of $E_{\rm int}$. 
This results in the shift of the positions of vortex cores of each component. 
Then, the last term of $E_{\rm int}$ can be interpreted effectively 
as the interaction energy between spins in the Ising model \cite{atten2}. 
Spin-up components correspond to the density peaks 
of $\Psi_{1}$ at the vortex cores of $\Psi_{2}$, 
and vice versa for spin-down components. 
When the coefficient $1 - \delta$ is positive, 
the interaction between spins is anti-ferromagnetic, 
which makes a square lattice stabilized (because the triangular lattice could be frustrated). 
Anti-ferromagnetic nature is made more pronounced as the interlocking of vortex 
lattices becomes stronger with $\delta$ and the number of vortices increase 
with $\bar{\Omega}$. 

\begin{figure}[btp]
\includegraphics[height=0.28\textheight]{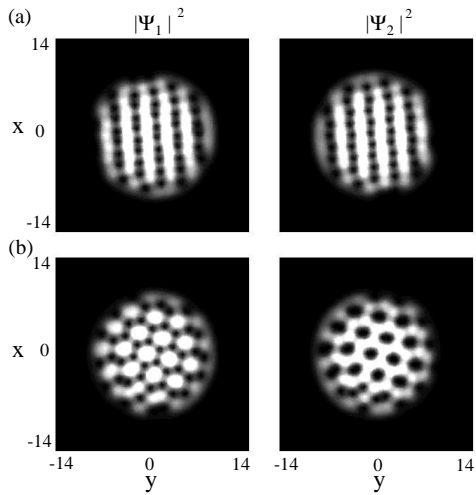}
\caption{The density profiles of the condensates $\Psi_{1}$ and $\Psi_{2}$ 
for $\bar{\Omega}=0.7$ and $\delta=1$. The profiles (a) and (b) are obtained by 
the numerical simulation starting from different trial wave functions. 
The energy deference between these states is 
$\Delta E \sim 10^{-5}$.}
\label{vordouble}
\end{figure}
As $\delta$ exceeds unity, the system enters a ferromagnetic phase. 
Then, the condensates undergo phase separation to spontaneously form domains 
having the same spin component \cite{atten2}. Concurrently, 
vortices of the same component begin to overlap at $\delta \sim 1$. 
Figure \ref{vordouble} shows typical solutions at $\delta = 1$. 
In Fig. \ref{vordouble}(a), vortices of each component are overlapped 
in lines, and each condensate density forms a stripe pattern \cite{Engels}. 
While this structure can also be derived if one assumes a perfect lattice \cite{Mueller}, 
we find that its energy is nearly degenerate with that of Fig. \ref{vordouble}(b). 
For the same parameters, we can obtain another equilibrium state called ``double-core vortex lattice" 
in Fig. \ref{phasedia}, where a vortex lattice of component 2 is made by pairs of vortices 
with the same circulation; vortices in component 1 surround those pairs. 
Therefore, various metastable structures will appear in this parameter region.

\begin{figure}[btp]
\includegraphics[height=0.46\textheight]{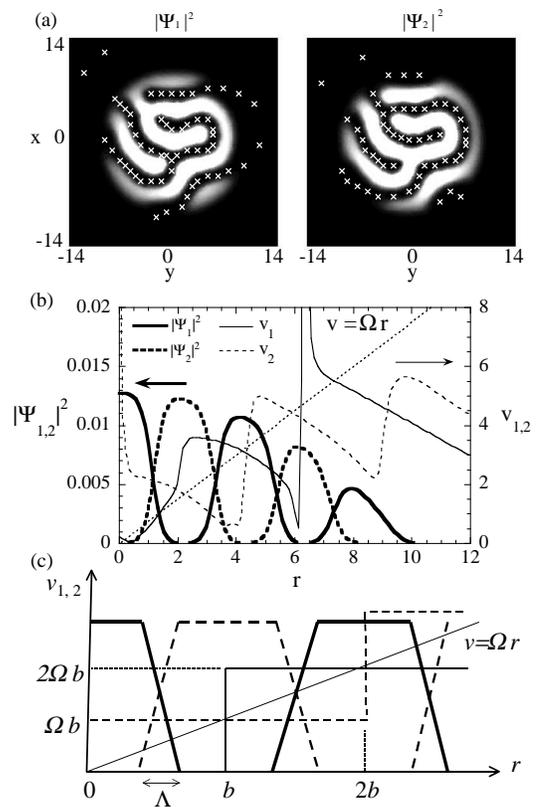}
\caption{(a) The density profiles of the condensates for $\bar{\Omega}=0.75$ and 
$\delta=1.1$. The vortex sheets are made up of chains of singly quantized vortices 
whose positions are marked by $\times$. 
(b) The density profile of $\Psi_{1}$ and $\Psi_{2}$ in the radial component, 
and the corresponding velocity profile.
The velocity of solid-body rotation is shown by the dotted line. 
(c) Schematic illustration of the model of the vortex sheet state. }
\label{vorsheet}
\end{figure}

For strongly phase-separated region $\delta>1$, 
the density peaks of the same spin component, 
at which the other-component vortices are located, 
merge further, resulting in the formation of vortex sheets; 
a typical example is shown in Fig. \ref{vorsheet}(a). 
Singly quantized vortices line up in sheets, and the sheets of component 1 and 2 
are interwoven alternately from the center to the outward. 
Figure \ref{vorsheet}(b) shows that each superfluid velocity $v_{i}$ $(i=1,2)$ jumps at the 
vortex sheet, following approximately the velocity of solid-body rotation. 
In the region $\delta>1$, even though the value of $\delta$ is changed, the total density 
is fixed by the Thomas-Fermi distribution with $\delta = 1$.

The stationary vortex sheet has been observed in rotating superfluid $^{3}$He-A \cite{sheethe}, 
where vortices are bounded to a topologically stable domain wall soliton across which 
the unit vector representing the direction of the orbital angular momentum 
of the Cooper-pair faces oppositely. 
As discussed in superfluid $^{3}$He-A, the equilibrium distance between 
the sheets of component 1 and component 2 is determined by the competition 
between the surface tension $\sigma$ 
of the domain wall and the kinetic energy of the superflow \cite{sheethe}. 
In order to estimate that distance in two-component BECs, 
we consider a simple model shown in Fig. \ref{vorsheet}(c). 
In this model, each velocity is assumed to be constant  
between the vortex sheets of the {\it same} component; 
the value of $v_{i}$ increases by $2 \bar{\Omega} b$ across every sheet. 
The total density $\rho_{T}=|\Psi_{1}|^{2}+|\Psi_{2}|^{2}$ is constant, 
and the domain boundary with the penetration depth $\Lambda(< b)$ is 
approximated by the linear profile. 
We calculate the free energy $F=E-\Omega L_{z}$ in the range $0<r<2b$, 
the sheet distance $b$ being determined so as to minimize $F$ per unit area. 
Firstly, the penetration depth $\Lambda$ is determined by minimizing the 
surface tension $\sigma$ of a single domain wall with respect to $\Lambda$ \cite{Tim}, 
which is the sum of the quantum pressure energy 
$E_{\rm qp}/2 \pi b \approx (\rho_{T}/2 \Lambda)$ and 
the interaction energy in the overlap region 
$E_{\rm int}/2 \pi b =(\delta-1) \Lambda \rho_{T}^{2}/12$; 
we thus obtain $\Lambda_{p} = \sqrt{6/C (\delta-1)\rho_{T}}$. 
Then, the surface tension is written as $\sigma = \rho_{T}^{3/2} \sqrt{C (\delta-1)/6}$.
Secondly, the flow energy per area (in rotating frame) is given by 
$(1/4\pi b^{2})\int_{0}^{2b} d^{2}r \sum_{i} \rho _{i} (v_{i}-\bar{\Omega} r)^{2}/2 = 
29 \rho_{T} \bar{\Omega}^{2} b^{2}/768$, where $b \ll \Lambda_{p}$ is assumed. 
Thus, the free energy per unit area is written as
\begin{equation}
\frac{F}{4 \pi b^{2}} = \frac{29 \rho_{T} \bar{\Omega}^{2} b^{2}}{768}
+ \frac{2 \sigma}{b}+\frac{C \rho_{T}^{2}}{2}. 
\end{equation}
Minimizing this energy with respect to $b$, one obtains $b=(768 \sigma/29 \bar{\Omega}^{2} \rho_{T})^{1/3}
=(768 /29 \bar{\Omega}^{2} \Lambda_{p})^{1/3}$.
By using the Thomas-Fermi density at $r=0$ with $\delta=1$
as the value of $\rho_{T}$, the sheet spacing 
$b \propto (\delta-1)^{1/6} (1-\bar{\Omega}^{2})^{1/12}/\bar{\Omega}^{2/3}$ is consistent 
with that of numerical solutions; for example, for parameters used in Fig. \ref{vorsheet} 
we obtain $b=3.09b_{\rm ho}$. 
The vortex sheet is expected in the region $b>\Lambda_{p}$ and $2b<R_{\rm TF}=\sqrt{2/\sqrt{\pi \alpha}}$, 
as shown in the shaded region in Fig. \ref{phasedia}. 
When $2b>R_{\rm TF}$, i.e., $b$ becomes comparable with the condensate size, 
the clear structure of sheets vanishes. 

Finally, we comment on what structure could be observed in actual experimental conditions. 
The two-component BECs realized in JILA is a mixture of the states 
$|1,-1 \rangle$ and $|2,1 \rangle$ of $^{87}$Rb \cite{Hall}. 
This mixture has the scattering lengths which have the same order, 
$a_{11} : a_{22} : a_{12} = 1:0.94:0.97$, i.e., $\delta \sim 1$ for equal particle numbers. 
Therefore, lattices with partially overlapping vortices of the same component, 
as shown in Fig. \ref{vordouble}, are expected to be observed. 
The MIT group has observed the phase separation of a mixture of $^{23}$Na BECs with $|1,0 \rangle$ 
and $|1,1 \rangle$ state \cite{Miesner}, 
which has $\delta \geq 1$ ($a_{11} : a_{22} : a_{12} = 1:1.035:1.035$).
In addition, a mixture of $^{41}$K and $^{87}$Rb BECs reported recently \cite{Modugno} 
lies deeply in a phase-separate region. 
The vortex sheets should therefore be observed at high rotation frequencies. 

In conclusion, we reveal a rich phase diagram of vortex states in rotating two-component BECs,
even in the restricted parameter space $( \delta, \bar{\Omega})$. 
The structure of a vortex lattice is shown to depend sensitively on 
the rotation frequency. In particular, we find new phenomena such as double-core 
vortex lattices and interwoven serpentine vortex sheets. 
Use of different atomic masses and intracomponent interactions will realize 
a coexistence system of vortices with different vortex-core sizes; 
such a situation may change the lattice structure shown in this Letter. 
We plan to study a more detailed phase diagram 
and effects of the internal Josephson coupling on the vortex states.

K.K. and M.T. acknowledge support by a Grant-in-Aid for Scientific Reserach (Grants No. 1505955 and No. 15341022) by the Japan Society for the Promotion of Science. 
M.U. acknowledges  support by a Grant-in-Aid for Scientific Research (Grant No.14740181) by the Ministry of Education, Culture, Sports, Science and Technology of Japan, and a CREST program by Japan Science and Technology Corporation (JST). 


\end{document}